\documentclass{article}
\usepackage{PRIMEarxiv}
\usepackage[utf8]{inputenc} % allow utf-8 input
\usepackage[T1]{fontenc}    % use 8-bit T1 fonts
\usepackage{hyperref}       % hyperlinks
\usepackage{url}            % simple URL typesetting
\usepackage{booktabs}       % professional-quality tables
\usepackage{amsfonts}       % blackboard math symbols
\usepackage{nicefrac}       % compact symbols for 1/2, etc.
\usepackage{microtype}      % microtypography
\usepackage{lipsum}
\usepackage{fancyhdr}       % header
\usepackage{graphicx}       % graphics
\graphicspath{{media/}}     % organize your images and other figures under 
\usepackage{booktabs}
\usepackage{multirow}
\usepackage{amsmath}
\title{Development of MR spectral analysis method robust against static magnetic field inhomogeneity}
\author{
  Shuki Maruyama \\
  Imaging Modality Group \\
  Advanced Technology Research Department \\
  Research and Development Center\\
Canon Medical Systems Corporation \\
	\texttt{shuki1.maruyama@medical.canon}
   \And
  Hidenori ~Takeshima \\
  Advanced Technology Research Department \\
  Research and Development Center \\
  Canon Medical Systems Corporation\\
}

\begin{document}
\maketitle

\begin{abstract}
	\textbf{Purpose:} To develop a method that enhances the accuracy of spectral analysis in the presence of static magnetic field ($\mathrm{B_0}$) inhomogeneity.\\[0.5mm]
\textbf{Methods:} The authors proposed a new spectral analysis method utilizing a deep learning model trained on modeled spectra that consistently represent the spectral variations induced by $\mathrm{B_0}$ inhomogeneity. These modeled spectra were generated from the $\mathrm{B_0}$ map and metabolite ratios of the healthy human brain. The $\mathrm{B_0}$ map was divided into a patch size of subregions, and the separately estimated metabolites and baseline components were averaged and then integrated. The quality of the modeled spectra was visually and quantitatively evaluated against the measured spectra. The analysis models were trained using measured, simulated, and modeled spectra. The performance of the proposed method was assessed using mean squared errors (MSEs) of metabolite ratios. The mean absolute percentage errors (MAPEs) of the metabolite ratios were also compared to LCModel when analyzing the phantom spectra acquired under two types of $\mathrm{B_0}$ inhomogeneity.\\[0.5mm]
\textbf{Results:} The modeled spectra exhibited broadened and narrowed spectral peaks depending on the $\mathrm{B_0}$ inhomogeneity and were quantitatively close to the measured spectra. The analysis model trained using measured spectra with modeled spectra improved MSEs by 49.89\% compared to that trained using measured spectra alone, and by 26.66\% compared to that trained using measured spectra with simulated spectra. The performance improved as the number of modeled spectra increased from 0 to 1,000. This model showed significantly lower MAPEs than LCModel under both types of $\mathrm{B_0}$ inhomogeneity.\\[0.5mm]
\textbf{Conclusion:} A new spectral analysis-trained deep learning model using the modeled spectra was developed. The results suggest that the proposed method has the potential to improve the accuracy of spectral analysis by increasing the training samples of spectra.
\end{abstract}

% keywords can be removed
% \keywords{First keyword \and Second keyword \and More}

\section{Introduction}
This study focuses on enhancing the accuracy of spectral analysis in magnetic resonance spectroscopy (MRS) for reliable metabolite quantification in the presence of static magnetic field ($\mathrm{B_0}$) inhomogeneity. The accuracy of spectral analysis in MRS is often limited by factors such as subject motion, low signal-to-noise ratio (SNR), poor water suppression, and $\mathrm{B_0}$ inhomogeneity\cite{Wilson2019}. $\mathrm{B_0}$ inhomogeneity causes peak broadening and distortion, hindering accurate metabolite quantification\cite{Juchem2021}. This can result in overlapping spectral peaks or loss of low-concentration metabolite peaks within the noise\cite{Tomiyasu2022}.\par
Conventional spectral analysis methods include peak integration, peak fitting, and linear combination modeling (LCM)\cite{Near2021}. While peak integration is simple, it lacks robustness in noisy or complex spectra\cite{NEAR201449}. Peak fitting\cite{kelm} improves resolution by modeling peak shapes but requires detailed prior knowledge, which is often unavailable, particularly for overlapping metabolite peaks in brain spectra.\par
LCM-based methods, such as LCModel\cite{Provencher1993}\cite{Provencher2001}, remain the most widely used for spectral analysis in MRS\cite{Near2021}. These methods fit the measured spectra using the weighted sums of the reference spectra, called basis sets, which are obtained from numerical simulations or phantom measurements. Several tools, including TARQUIN\cite{Wilson2011}, QUEST/JMRUI\cite{Stefan2009}, and Osprey\cite{Oeltzschner2020}, have extended the LCM framework by providing enhanced modeling capabilities and greater flexibility in both the time and frequency domains.\par
This study aims to develop a new method to improve the accuracy of spectral analysis in the presence of $\mathrm{B_0}$ inhomogeneity. One major limitation of LCM-based methods is their sensitivity to $\mathrm{B_0}$ inhomogeneity, which can result in degraded fitting performance\cite{Provencher2001}. Recent advances in deep learning have shown promise in spectral analysis\cite{Hatami2018}, particularly under low SNR and poor shimming conditions\cite{Lee2019}\cite{Lee2020}\cite{Shamaei2023}\cite{Chen2024}. These methods require large and realistic training datasets that are difficult to obtain from in vivo measurements. Thus, simulated spectra were employed for dataset generation. However, simulated spectra often fail to accurately capture the spectral variability induced by $\mathrm{B_0}$ inhomogeneity in actual measurements. Discrepancies between simulated and measured spectra may cause errors in metabolite quantification\cite{Shamaei2023}. When applying modeled spectra to the training of deep learning models, the closeness to the measured spectra is an essential factor\cite{Shrivastava2017}\cite{Baradad2021}.

\section{Materials and Methods}
\subsection{Overview}
The authors proposed a new spectral analysis method utilizing a deep learning model trained on modeled spectra that consistently represent the spectral variations induced by the $\mathrm{B_0}$ inhomogeneity (Figure 1). 

\begin{figure}[h]
    \centering
    \includegraphics[width=1\linewidth]{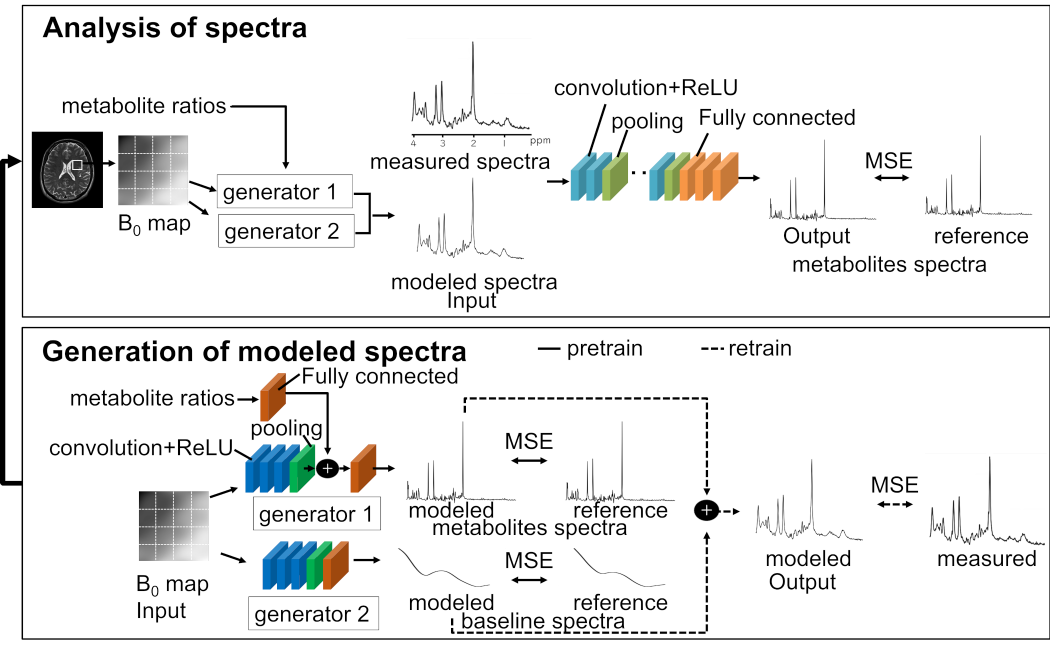}
    \caption{Overview of proposed method. The generative models are trained to generate modeled spectra by integrating $\mathrm{B_0}$ map and metabolite ratios. The modeled spectra are compared to the measured spectra for visual and quantitative evaluations. The proposed method trains the analysis model using measured spectra along with the modeled spectra. The metabolite quantification is evaluated on healthy human brain and phantom spectra under variations of $\mathrm{B_0}$ inhomogeneity.
$\mathrm{B_0}$, static magnetic field; MSE, mean squared error; ReLU, rectified linear unit.
}
    \label{fig1:enter-label}
\end{figure}

Generative models were trained to generate the modeled spectra by integrating the $\mathrm{B_0}$ map and metabolite ratios. The modeled spectra were compared with the measured spectra for visual and quantitative evaluation. The proposed method trained the analysis models using the measured and modeled spectra. Metabolite quantification was evaluated using the spectra of healthy human brains and phantom spectra under two types of $\mathrm{B_0}$ inhomogeneity. The total creatinine (tCr: creatine + phosphocreatine) was used as the reference metabolite for quantification.

\subsection{Spectral Generative Models}
A new spectral generative model, which models $\mathrm{B_0}$ inhomogeneity, was constructed to increase the training samples of spectra for spectral analysis (Figure 1, bottom). The model comprises two-dimensional convolutional neural networks (CNNs), referred to as generators 1 and 2. Generator 1 extracts features from the $\mathrm{B_0}$ map and metabolite ratios as inputs and generates the modeled metabolites spectra. Generator 2 extracts features from the $\mathrm{B_0}$ map as inputs and generates the modeled baseline spectra.\par 
A $\mathrm{B_0}$ map and 15 metabolite ratios were used as inputs for the two generator networks. Generator 1 (788,735 parameters) for the modeled metabolites spectra contained three convolutional layers, an averaging pooling layer, and two fully connected layers. Generator 2 (110,001 parameters) for the modeled baseline spectra contained three convolutional layers, an averaging pooling layer, and a fully connected layer. All convolutional layers had a kernel size of $\mathrm{3\times3}$ and a stride of 1. A rectified linear unit (ReLU) activation function\cite{Nair2010RectifiedLU} was used for each convolutional layer. The $\mathrm{B_0}$ map was cropped to the volume of interest (VOI) of the measured spectra and resampled to a size of $\mathrm{128\times128}$ pixels. The intensities of the $\mathrm{B_0}$ map, reference metabolites, and baseline spectra were normalized to the range [0,1]. The number of spectral points was 379, corresponding to the chemical shift range of 0.2–4.1 ppm.

\subsection{Training of Spectral Generative Models}The generative model was trained to produce modeled spectra that reflect the spectral variability induced by $\mathrm{B_0}$ inhomogeneity. The model was trained under five conditions of the training method (Figure 2).

\begin{figure}[h]
    \centering
    \includegraphics[width=1\linewidth]{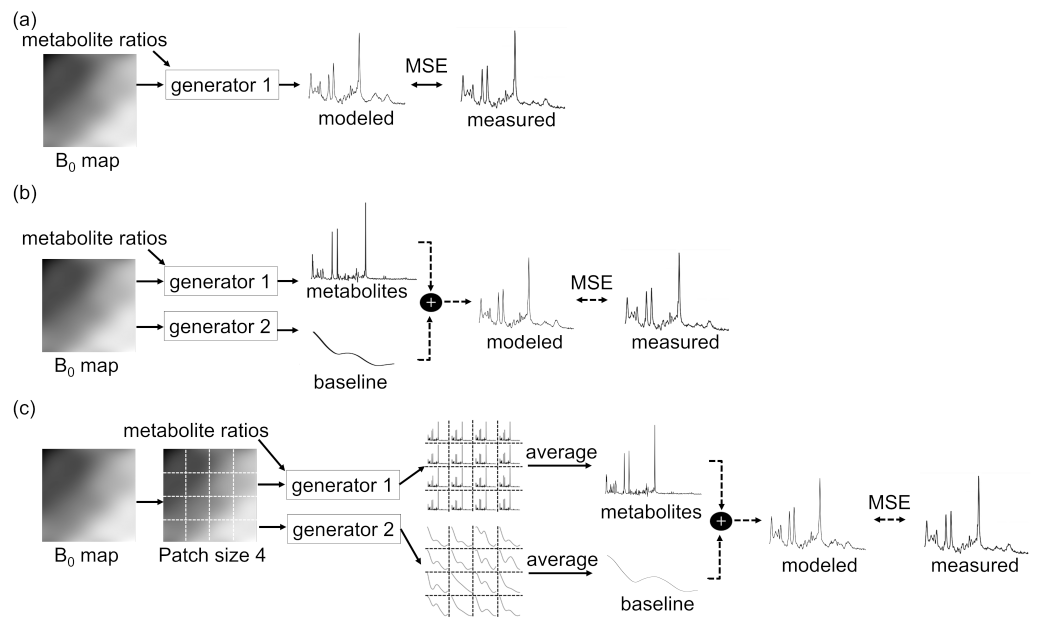}
    \caption{Training methods of proposed generative model. The model is trained to estimate the measured spectra using one-step (a) and two-step methods. The two-step method trains the model to generate spectra of metabolites and baseline components. These spectra are concatenated and used to retrain the model, with the measured spectra provided as the target (b). The $\mathrm{B_0}$ map is divided into patches of subregions before input, and the model generates spectra for each subregion; these spectra are then averaged (c).
}
    \label{fig2:enter-label}
\end{figure}

In the first setting, the model was trained to directly estimate the measured spectra in a single step (Figure 2a). The second setting involved a two-step estimation of the measured metabolites and baseline spectra (Figure 2b). In the first step, the model was trained to generate the spectra of metabolites and baseline components, with the basis set and reference baseline provided as targets. In the second step, these spectra were concatenated and used to retrain the model with the measured spectra provided as the target. The remaining three conditions were used to explore the effect of the average output spectra, in which the input $\mathrm{B_0}$ map was divided into patches (Figure 2c). Patch sizes of two, four, and eight were used in this study. The measured spectrum represented the average over the VOI. The $\mathrm{B_0}$ map was divided into a patch of subregions before input, with each subregion associated with a distinct $\mathrm{B_0}$ profile. The model generated spectra for each subregion, and the spectra were averaged. The model was pretrained to estimate the metabolites spectra of the basis set and baseline spectra, followed by retraining using the target measured spectra.

The generative models were trained with an Adam optimizer\cite{Kingma2015} for 500 epochs, with a learning rate for all networks set to 0.0001, an exponential decay rate for the first moment estimates of $\beta_1$ = 0.9, an exponential decay rate for the second moment estimates of $\beta_2$ = 0.999, and a mini-batch size of 16. The loss function was mean squared errors (MSE). These hyperparameters were selected based on preliminary experiments. 

\subsection{Evaluation of Spectral Generative Models}
The modeled spectra were visually and quantitatively evaluated to ensure that they were close to the measured spectra used for training the analysis model. The MSE was calculated against the measured spectra to quantitatively evaluate the modeled spectra. 

\subsection{Spectral Analysis Models}
To estimate the metabolite ratios from the spectra, a spectral analysis model was constructed following the methodology proposed by Lee et al\cite{Lee2020}(Figure 1, top). The model comprises a one-dimensional CNN. The CNN extracts features from the measured or modeled spectrum as input and produces metabolites spectra.\par 
The CNN (121,327 parameters) contained eight convolutional layers, four max pooling layers, and three fully connected layers. The convolutional layers had a kernel size of 3 and a stride of 1. A ReLU activation function followed each convolutional layer. 

\subsection{Training of Spectral Analysis Models}
Spectral analysis models were trained to evaluate the effectiveness of the proposed method. They were trained under five training data conditions: using measured spectra alone, using conventionally simulated spectra\cite{Shamaei2023} alone, using modeled spectra alone, using measured spectra with simulated spectra, and using measured spectra with spectra modeled by the proposed generative model. The trained generative models were used to generate modeled spectra from two inputs: $\mathrm{B_0}$ maps randomly extracted from regions not containing the dura mater and metabolite ratios adopted from Shamaei et al\cite{Shamaei2023}.\par 
The analysis models were trained with an Adam optimizer for 100 epochs, with a learning rate for all networks set to 0.0001, an exponential decay rate for the first moment estimates of $\beta_1$ = 0.9, an exponential decay rate for the second moment estimates of $\beta_2$ = 0.999, and a mini-batch size of 4. The loss function was the MSE. These hyperparameters were selected based on preliminary experiments. 

\subsection{Evaluation of Spectral Analysis Models}
To evaluate the analysis models, the MSEs against the reference metabolites spectra were calculated for the five conditions described above. The number of modeled and simulated spectra for training the models was set to 1,000. The simulated spectra are available at \href{https://github.com/isi-nmr/Deep-MRS-Quantification}{https://github.com/isi-nmr/Deep-MRS-Quantification}. In addition, the effect of the number of the modeled spectra added to the training data on the performance of the analysis models was investigated. The MSEs against the reference metabolite ratios were compared when the number of modeled spectra was set to 100, 1,000, and 10,000.\par 
The spectra of a phantom mimicking the in vivo healthy human brain were analyzed to evaluate the accuracy of metabolite quantification under varying $\mathrm{B_0}$ inhomogeneity. Quantification accuracy was measured using mean absolute percentage errors (MAPEs) of the metabolite ratios and compared to LCModel. MAPEs were computed for total NAA (tNAA: N-acetylaspartate + N-acetylaspartylglutamate), total Cho (tCho: glycerophosphocholine + phosphocholine), myo-inositol (Ins), and Glx (glutamate-glutamine), with \%SD < 15\% in the LCModel analysis. 

\subsection{Volunteer Dataset}
A volunteer dataset was obtained from the dataset used by Maruyama et al\cite{Maruyama2024} to train and evaluate our models. This dataset included $\mathrm{B_0}$ maps and spectra measured from forty-six healthy human brains. This dataset was approved for use in this study by the Institutional Review Board, and informed consent was obtained from all participants before their participation. A total of 109 spectra from thirty participants were used for training, 32 spectra from seven participants for validation, and 33 spectra from nine participants for testing the proposed models.\par
The $\mathrm{B_0}$ maps were computed from image-based shimming data with two echo times ($\mathrm{TE_1}$ and $\mathrm{TE_2}$) using a dual-echo gradient sequence, the details of which were not described by Maruyama et al \cite{Maruyama2024}. The measurement parameters were TR = 400 ms, $\mathrm{TE_1}$ = 4.8 ms, $\mathrm{TE_2}$ = 7.2 ms, flip angle = 70°, voxel size = $\mathrm{3.4 \times 3.4 \times 2.0}$ mm, matrix = $\mathrm{64 \times 64}$, bandwidth = 977 Hz, FOV = $\mathrm{220 \times 220}$ mm, number of excitations (NEX) = 1, acquisition time = 26 s, and slices = 30. Since the spectra were measured from the frontal, parietal, and occipital lobes of the brain using a point-resolved spectroscopy (PRESS) sequence\cite{Bottomley1987} (TR/TE = 2000/34 ms) after $\mathrm{B_0}$ shimming, they included variability in $\mathrm{B_0}$ inhomogeneity. A detailed description of the measurement parameters of single-voxel spectroscopy is provided in a previous publication\cite{Maruyama2024}.\par
The spectra were reanalyzed for the purpose of this study. The measured spectra were prepared by post-processing the phase corrections of the raw spectra. The reference metabolites spectra were obtained from the basis set of the 3T PRESS sequence at TE = 35 ms provided by LCModel. The reference metabolite ratios and baseline spectra were obtained using LCModel (version 6.3-1N; Stephen Provencher, Inc., Oakville, Ontario, Canada) analysis of the measured spectra using the basis set. The basis set consisted of the spectra for the following metabolites: alanine, aspartate, creatine (Cr), phosphocreatine, gamma-aminobutyric acid (GABA), glutamine, glutamate, glycerophosphocholine, phosphocholine, glutathione, Ins, NAA, N-acetylaspartylglutamate, lactate, and taurine. 

\subsection{Phantom Dataset}
An in vitro phantom dataset was constructed to evaluate the accuracy of metabolite quantification by the proposed method under $\mathrm{B_0}$ inhomogeneity. The phantom spectra were measured seven times using the same parameters as those for the volunteer dataset. To represent two types of $\mathrm{B_0}$ inhomogeneity, VOIs were placed near center and the periphery of the phantom, corresponding to regions of low and high $\mathrm{B_0}$ inhomogeneity, respectively. The phantom contained 12.5 mM NAA, 10.0 mM Cr, 3.0 mM choline, 12.5 mM glutamate, 7.5 mM Ins, 5.0 mM lactate, and 2.0 mM GABA prepared in 500 mL of phosphate-buffered saline. The pH was adjusted to 7.2 with NaOH.

\subsection{Statistical Analysis}
Two-way analysis of variance (ANOVA) was performed on MAPEs with factors of the analysis methods (proposed method and LCModel) and metabolite ratios (Ins, tCho, tNAA, and Glx), followed by the Sidak multiple comparison test. A \textit{P}-value less than 0.05 indicated statistical significance. All analyses were performed using Prism 9 software (GraphPad Software, San Diego, CA, USA).

\section{Results}
\subsection{Spectral Generation}
Table 1 presents a comparison of the MSEs of the modeled spectra generated by the one-step and two-step methods using patch sizes of 1, 2, 4, and 8. The two-step method exhibited a lower MSE than the one-step method. The MSEs were minimized when the patch size was 4. 
\begin{table}[h]
	\caption{Comparison of MSEs of modeled spectra generated with one-step and two-step methods using patch sizes of 1, 2, 4, and 8.  MSE, mean squared error.}
	\centering
    \scalebox{1.1}[1.0]{
	\begin{tabular}{lcr}
		\toprule
		% \cmidrule(r){1-2}
		Training method    & Patch size     & MSE \\
		\midrule
		One-step     & 1  & 0.0252     \\ \midrule
		\multirow{4}{*}{Two-step}     & 1  &  0.0215      \\
		     & 2       &0.0209  \\
                 & 4       & 0.0205  \\
                & 8       & 0.0220  \\
		\bottomrule
	\end{tabular}}
	\label{tab:table}
\end{table}

Figure 3 shows a comparison of the modeled spectra generated using the one-step method (Figure 3a), two-step method (Figure 3b), two-step method using a patch size of four (Figure 3c), and the measured spectra (Figure 3d) obtained from the same participant. The modeled spectra generated using the two-step method more closely resembled the measured spectra than those generated using the one-step method, particularly at the peaks indicated by the black arrows. The modeled spectra generated using the one-step method contained more noise than those generated using the other methods. No significant visual differences were observed with the varying patch sizes.

\begin{figure}[h]
    \centering
    \includegraphics[width=0.5\linewidth]{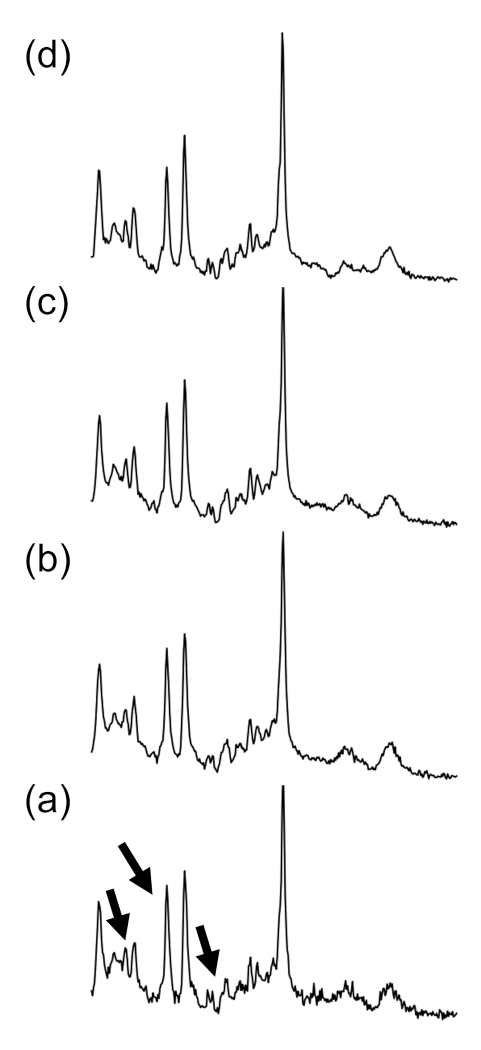}
    \caption{Comparison of modeled spectra from training methods. The modeled spectra were generated using the one-step method (a), the two-step method with a patch size of one (b), the two-step method with a patch size of four (c), and measured spectra (d), obtained from the same participant.
}
    \label{fig:enter-label}
\end{figure}

Figure 4 shows a comparison between the modeled spectra, $\mathrm{B_0}$ map, and the measured spectra obtained from the same participant in regions with lower (left) and higher (right) $\mathrm{B_0}$ inhomogeneity. The spectra generated by the proposed generative model were close to the measured spectra in both regions. Broader spectral peaks were observed in the region with higher $\mathrm{B_0}$ inhomogeneity, whereas narrower peaks were observed in the region with lower $\mathrm{B_0}$ inhomogeneity. The $\mathrm{B_0}$ inhomogeneity was increased by magnetic susceptibility variations near the cerebrospinal fluid. The water linewidths were 8.6 Hz and 9.9 Hz. The modeled spectra were generated using a two-step method with a patch size of 4. 

\begin{figure}[h]
    \centering
    \includegraphics[width=1\linewidth]{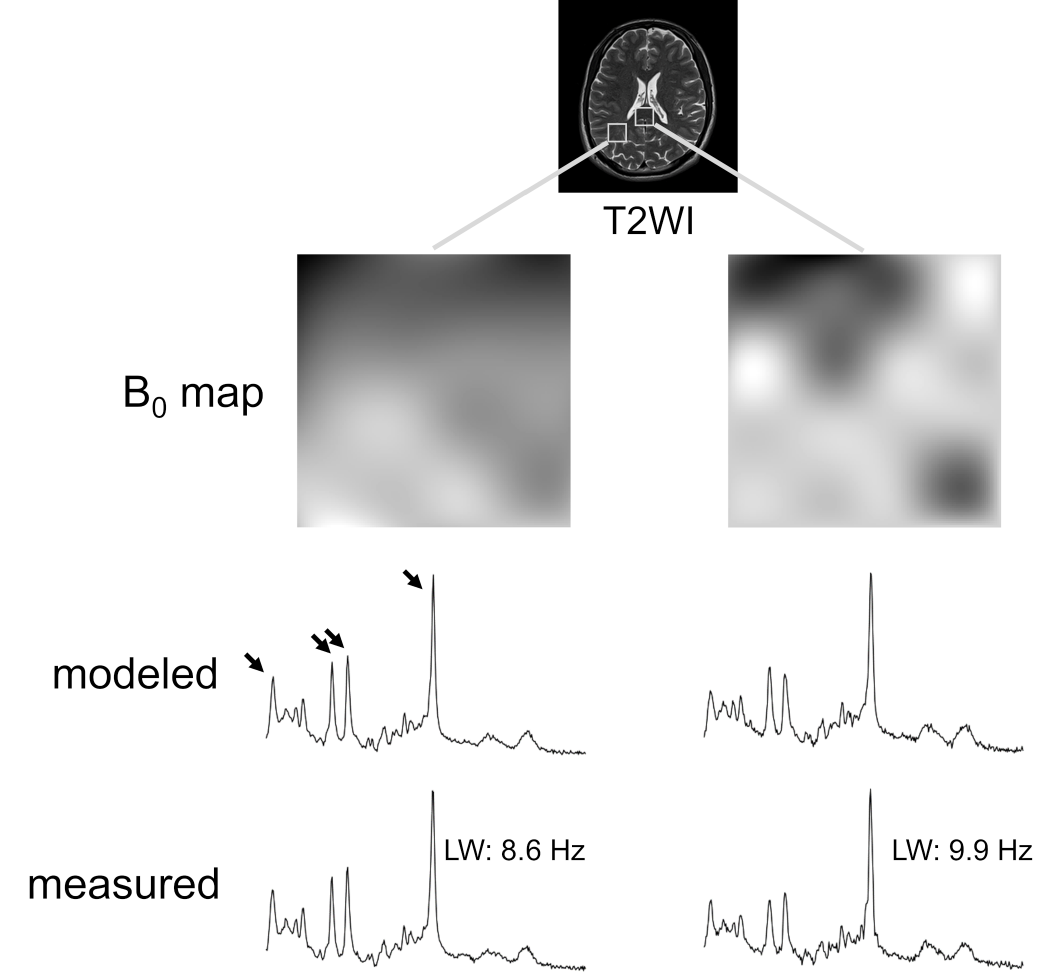}
    \caption{Comparison of modeled spectra generated from two types of $\mathrm{B_0}$ maps. The modeled spectra, $\mathrm{B_0}$ map, and measured spectra in regions with higher (right) and lower (left) $\mathrm{B_0}$ inhomogeneity, obtained from the same participant. The $\mathrm{2\times2\times2}$ $\mathrm{cm^3}$ volume of interest (white box) is superimposed on the T2-weighted image (T2WI).
LW, water linewidth.
}
    \label{fig:enter-label}
\end{figure}

\subsection{Spectral Analysis}
Table 2 compares the MSEs of the analysis models trained using the measured spectra alone, conventionally simulated spectra alone, modeled spectra alone, measured spectra with simulated spectra, and measured spectra with modeled spectra. The analysis model trained using measured spectra with modeled spectra resulted in the lowest MSE among all methods. Specifically, it achieved a 1350.10\% reduction in MSE compared to that trained using simulated spectra alone, a 400.43\% reduction compared to that trained using modeled spectra alone, a 49.89\% reduction compared to that trained using measured spectra alone, and a 26.66\% reduction compared to that trained using measured spectra with simulated spectra.\par

\begin{table}[h]
	\caption{Comparison of MSEs of proposed method trained using measured spectra alone, using simulated spectra alone, using modeled spectra alone, using measured spectra with simulated spectra, and using measured spectra with modeled spectra. MSE, mean squared error.}
	\centering
	\scalebox{1.1}[1.1]{\begin{tabular}{lc}
		\toprule
		\cmidrule(r){1-2}
		Training data     & MSE \\
		\midrule
		109 measured spectra     & 0.0109     \\
		109 simulated spectra   & 0.1055   \\
		109 modeled spectra       &0.0364  \\
        109 measured + 1,000 simulated spectra       & 0.0092 \\ 109 measured + 1,000 modeled spectra (proposed)       & 0.0073  \\
		\bottomrule
	\end{tabular}}
	\label{tab:table}
\end{table}

Figure 5 shows a comparison of the MSEs of the analysis model trained using the measured spectra alone, measured spectra with 100 modeled spectra, measured spectra with 1,000 modeled spectra, and measured spectra with 10,000 modeled spectra. The MSEs were lower for models trained using the measured spectra with the modeled spectra than for those trained using the measured spectra alone. Performance improved as the number of modeled spectra increased from 0 to 1,000. No further improvement was observed when the number of modeled spectra increased from 1,000 to 10,000.\par

\begin{figure}[h]
    \centering
    \includegraphics[width=1\linewidth]{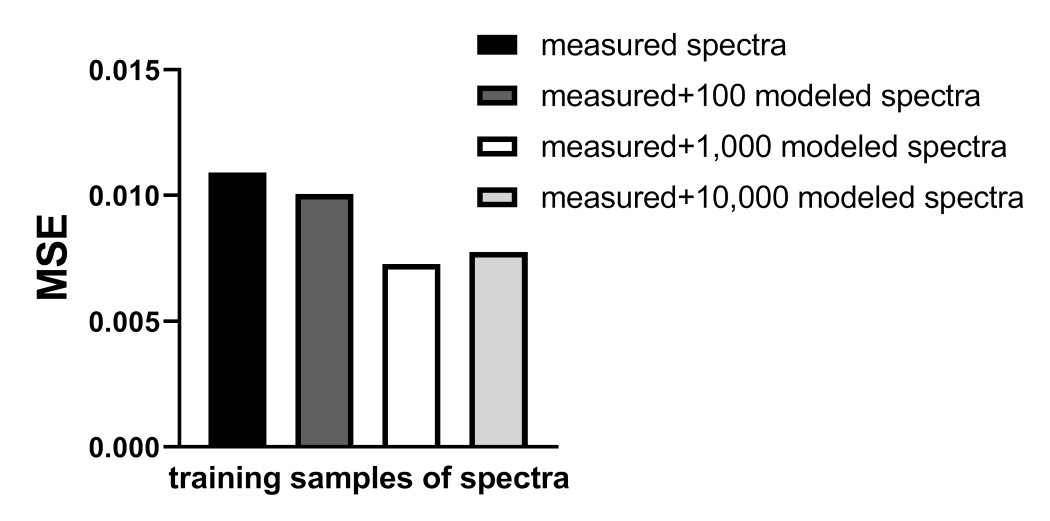}
    \caption{Comparison of MSEs of analysis model when increasing number of modeled spectra for training. The proposed method trained the analysis model using measured spectra alone, measured spectra with 100 modeled spectra, measured spectra with 1,000 modeled spectra, and measured spectra with 10,000 modeled spectra. 
MSE, mean squared error.
}
    \label{fig:enter-label}
\end{figure}

Figure 6 shows a comparison of the MAPEs of the proposed method and LCModel when the spectra were measured near the center (Figure 6a) and periphery (Figure 6b) of the phantom. The water linewidths were 8.0 ± 0.4 Hz (mean ± SD) at the near center and 10.9 ± 0.6 Hz (mean ± SD) at the periphery. At the near center, two-way ANOVA showed a significant main effect of the analysis method ($F_{(1,48)}$ = 256.5, \textit{P} < 0.0001), metabolite ratio ($F_{(3,48)}$ = 88.31, \textit{P} < 0.0001), and interaction ($F_{(3,48)}$ = 65.77, \textit{P} < 0.0001). The Sidak multiple comparison test revealed significant differences in Ins (\textit{P} < 0.0001), tCho (\textit{P} < 0.0001), and tNAA (\textit{P} = 0.0016). At the periphery, two-way ANOVA showed a significant main effect of the analysis method ($F_{(1,48)}$ = 170.50, \textit{P} < 0.0001), metabolite ratio ($F_{(3,48)}$ = 84.77, \textit{P} < 0.0001), and interaction ($F_{(3,48)}$ = 121.9, \textit{P} < 0.0001). The Sidak multiple comparison test revealed a significant difference in Ins (\textit{P} < 0.0001) and Glx (\textit{P} < 0.0001). 

\begin{figure}[h]
    \centering
    \includegraphics[width=1\linewidth]{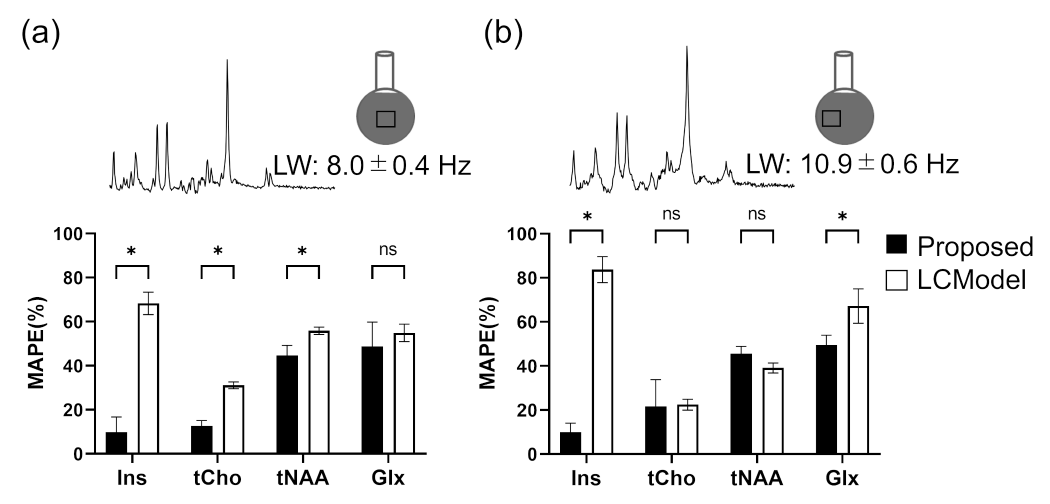}
    \caption{Comparison of MAPEs of proposed method and LCModel. The spectra were measured at the near center (a) and the periphery (b) of the phantom. The water linewidths were 8.0 ± 0.4 Hz (mean ± SD) at the near center and 10.9 ± 0.6 Hz (mean ± SD) at the periphery. * \textit{P} < 0.05 for comparison between the proposed method and LCModel using the Sidak multiple comparison test. The proposed method trained the analysis model using measured spectra with 1,000 modeled spectra.
MAPE, mean absolute percentage error; Ins, myo-inositol; tCho, total glycerophosphocholine + phosphocholine; tNAA, total N-acetylaspartate + N-acetylaspartylglutamate; Glx, glutamate-glutamine.
}
    \label{fig:enter-label}
\end{figure}

\section{Discussion}
The proposed method improved the quantification accuracy under varying $\mathrm{B_0}$ inhomogeneities, which is an important consideration for clinical applications. The results suggest that the proposed method has the potential to enhance metabolite quantification performance beyond that of LCModel. To the best of the authors’ knowledge, this report is the first to demonstrate the improvement of spectral analysis by explicitly modeling $\mathrm{B_0}$ inhomogeneity.\par
As shown in Table 2, the proposed method achieved the lowest MSEs among the analysis models trained using only the measured spectra, suggesting that the modeled spectra are useful for increasing the volume of training data for the spectral analysis model. The MSEs of the analysis model trained using both the measured and simulated spectra were reduced. This finding may be consistent with that of previous studies regarding spectral analysis trained using simulated spectra alone\cite{Shamaei2023}.\par
The analysis model trained using only the modeled spectra yielded a higher MSE than the model trained using the measured spectra, indicating that the modeled spectra may not completely represent the features of the measured spectra. However, these findings suggest that the discrepancy between the modeled and measured spectra is smaller than that between the simulated and measured spectra. This improvement is likely attributable to the fact that the modeled spectra were generated under conditions designed to closely approximate actual measurement conditions, including spatially varying $\mathrm{B_0}$ maps and region-wise spectral averaging.\par
The quantitative quality of the modeled spectra was improved by combining the two-step and patch-division methods during training. As shown in Table 1, the two-step method yielded lower MSEs than the one-step method, suggesting that separately estimating the metabolite and baseline components leads to more accurate spectral modeling. The MSEs were reduced when the input $\mathrm{B_0}$ maps were divided into two and four patches, indicating that averaging the spectra of patches affected by the local $\mathrm{B_0}$, as in the actual measurement method, was effective. The MSE increased when the patch size was increased to eight, possibly because of the reduced differences in spectral features across neighboring patches during training.\par 
As shown in Figures 3–4, the proposed generative model successfully generated modeled spectra close to the measured spectra, which is consistent with the quantitative findings. Furthermore, the modeled spectra exhibited broadened and narrowed spectral peaks depending on the types of $\mathrm{B_0}$ inhomogeneity (Figure 4). This finding indicates that the proposed generative model effectively modeled spectral variations induced by $\mathrm{B_0}$ inhomogeneity. Based on the visual and quantitative results, the modeled spectra were close to the measured spectra and contained as much diversity as possible by changing the inputs of the $\mathrm{B_0}$ map and metabolite ratios.\par
As shown in Figure 5, increasing the number of modeled spectra for training improved the performance of the proposed method. The reduction in MSEs observed when increasing the number of modeled spectra from 0 to 1,000 suggests that the additional modeled spectra help the analysis model generalize better by exposing it to a broader range of spectral variations, particularly those induced by $\mathrm{B_0}$ inhomogeneity. The absence of further improvement when the number of modeled spectra was increased from 1,000 to 10,000 implies a saturation point. These findings highlight the importance of balancing the measured and modeled spectra in training data.\par
As shown in Figure 6, the proposed method achieved significantly lower MAPEs than LCModel when analyzing the measured spectra of the phantom under the two types of $\mathrm{B_0}$ variation. Improvements were observed for Ins, tCho, and tNAA in the central region, and for Ins and Glx in the peripheral region. The peaks of Ins\cite{kaiser} and Glx\cite{Yahya2008} could be contaminated by peaks of other metabolites with higher $\mathrm{B_0}$ inhomogeneity. Thus, these findings suggest that the proposed method can improve accuracy and robustness under varying $\mathrm{B_0}$ conditions.

\section{Conclusion}
A new spectral analysis-trained deep learning model using the modeled spectra was developed. The modeled spectra consistently represented the spectral variations induced by $\mathrm{B_0}$ inhomogeneity. The results suggest that the proposed method has the potential to improve the accuracy of spectral analysis by increasing the training samples of spectra.

\section{Conflicts of Interest}
All authors are employees of Canon Medical Systems Corporation.

%Bibliography
\bibliographystyle{unsrt}  
\bibliography{references}

\end{document}